\documentclass [lettersize,journal]{IEEEtran}
\usepackage{amsmath,amssymb,amscd}
\usepackage{algorithm}
\usepackage{array}
\usepackage[caption=false,font=normalsize,labelfont=sf,textfont=sf]{subfig}
\usepackage{textcomp}
\usepackage{stfloats}
\usepackage{url}
\usepackage{verbatim}
\usepackage{graphicx}
 \usepackage{cite}

\usepackage{algpseudocode}
\usepackage{multirow}
\usepackage{amsmath,systeme}
\usepackage{mathtools}
\usepackage{xcolor}
\usepackage{tikz}
\usepackage{orcidlink}

\newcommand{\RNum}[1]{\uppercase\expandafter{\romannumeral #1\relax}}

\newtheorem{lemma}{Lemma}
\newtheorem{remark}{Remark}

\begin{document}

\title{Energy-Efficient Secure Offloading System Designed via UAV-Mounted Intelligent Reflecting Surface for Resilience Enhancement}

\author{Doyoung Kim\orcidlink{0009-0008-1098-9111},~\IEEEmembership{Student Member,~IEEE}, Seongah Jeong\orcidlink{0000-0002-9737-0432},~\IEEEmembership{Member,~IEEE}, and Jinkyu Kang\orcidlink{0000-0001-6111-937X},~\IEEEmembership{Member,~IEEE}
\thanks{This work was supported by the National Research Foundation of Korea(NRF) grant funded by the Korea government(MSIT) (No. 2023R1A2C2005507). 
This research was supported by the MSIT(Ministry of Science and ICT), Korea, under the ITRC(Information Technology Research Center) support program(IITP-2020-0-01787) supervised by the IITP(Institute of Information \& Communications Technology Planning \& Evaluation). 
This work was supported by the National Research Foundation of Korea(NRF) grant funded by the Korea government(MSIT) (No. 2021R1F1A1050734).
(\textit{Corresponding author}: Seongah Jeong and Jinkyu Kang.)}
\thanks{Doyoung Kim is with the School of Electronics Engineering, Kyungpook National University, Daegu 14566, South Korea (Email: singha5036 @knu.ac.kr).}
\thanks{Seongah Jeong is with the School of Electronics Engineering, Kyungpook National University, Daegu 14566, South Korea (Email: seongah@knu.ac.kr).}
\thanks{Jinkyu Kang is with the Department of Information and Communications Engineering, Myongji University, Gyeonggi-do 17058, South Korea (Email: jkkang@mju.ac.kr).}}

\maketitle
\begin{abstract}
With increasing interest in mmWave and THz communication systems, an unmanned aerial vehicle (UAV)-mounted intelligent reflecting surface (IRS) has been suggested as a key enabling technology to establish robust line-of-sight (LoS) connections with ground nodes owing to their free mobility and high altitude, especially for emergency and disaster response. This paper investigates a secure offloading system, where the UAV-mounted IRS assists the offloading procedures between ground users and an access point (AP) acting as an edge cloud. In this system, the users except the intended recipients in the offloading process are considered as potential eavesdroppers. The system aims to achieve the minimum total energy consumption of battery-limited ground user devices under constraints for secure offloading accomplishment and operability of UAV-mounted IRS, which is done by optimizing the transmit power of ground user devices, the trajectory and phase shift matrix of UAV-mounted IRS, and the offloading ratio between local execution and edge computing based on the successive convex approximation (SCA) algorithms. Numerical results show that the proposed algorithm can provide the considerable energy savings compared with local execution and partial optimizations.
\end{abstract}


\section{Introduction}\label{sec:intro}
\IEEEPARstart{I}{ntelligent} reflecting surface (IRS) has become a key enabling technology to enhance the received power and mitigate interference in ground-to-ground or air-to-ground communication with the benefit of its cost-effectiveness and lightweight \cite{Wu21JSAC, Wu21TWC, N20Arxiv, Li20WCL,Yang20TVT, Huang19TWC, Zhang19globecom, Shafique21TCOM, lu2020enabling, Long20Arxiv}. An IRS comprises numerous passive reflecting elements, each capable of adjusting the reflection coefficients, including the amplitude and phase shift, upon reflecting electromagnetic waves. With IRS, it is possible to reconfigure wireless channels without deploying costly active base stations (BSs) or relays. This is highly desirable for future 6G ecosystems, such as mmWave and THz communication systems with the high frequencies, where severe pathloss and blockage exist \cite{Wu21TWC, N20Arxiv}. Depending on the deployment location, various types of IRS configurations are considered, such as terrestrial IRS \cite{Li20WCL, Yang20TVT} attached to the walls or the surface of ground automobiles. The terrestrial IRS deployed between transceivers, however, has limitations for achieving the desirable virtual links in the presence of structures such as skyscrapers, lamp posts and billboards. Also, the terrestrial wireless infrastructures including terrestrial IRS are susceptible to become aging and damaged by the climate events such as extreme temperature, hurricanes, floods and so on. To this end, the unmanned aerial vehicle (UAV)-mounted IRS has been developed, leveraging free mobility and high altitude so as to increase the likelihood of establishing strong line-of-sight (LoS) links between transceivers \cite{Zhang19globecom, Shafique21TCOM, lu2020enabling, Long20Arxiv}.

Several works on UAV-mounted IRS for communication systems have been considered owing to its advantages \cite{Zhang19globecom, Shafique21TCOM, lu2020enabling, Long20Arxiv}. In \cite{Zhang19globecom}, an active self-steering mechanism for UAV-mounted IRS is developed via reinforcement learning (RL) strategy derived from Q-learning to maximize the downlink mmWave communication capacity. The authors in \cite{Shafique21TCOM} propose the use of UAV-mounted IRS for a full-duplex relaying system to provide an extra degree of freedom and analyze end-to-end outage probability, ergodic capacity, and energy efficiency. In \cite{lu2020enabling}, the UAV-mounted IRS placement, phase shift and transmit beamforming at BS are jointly optimized via the max--min problem regarding a signal-to-noise ratio (SNR) within a specific service zone. In contrast to the work in \cite{Zhang19globecom, Shafique21TCOM, lu2020enabling} that focus on utilizing UAV-mounted IRS for enhancing the primary end-to-end communication capacity, the authors in \cite{Long20Arxiv} employ the UAV-mounted IRS for secure networks under the existence of eavesdroppers, and propose the joint optimization of UAV-mounted IRS trajectory and passive beamforming to maximize the secure energy efficiency, where the UAV-mounted IRS acts as the relay between BS and ground users.

Recently, the IRS has also been studied for secure offloading and communication systems \cite{Li21IoTJ, Mao22TVT, xu2022computation}. By adjusting the beam direction and establishing the good channel quality with the IRS, the security issue and eavesdropping problem caused by the open attributes of the wireless medium can be alleviated. The authors in \cite{Li21IoTJ} investigate the secrecy improvement of mobile edge computing (MEC) systems with the aid of IRS for Internet-of-Things (IoT) devices. \cite{Mao22TVT} also proposes an IRS-assisted secure MEC system to maximize the minimum computation energy efficiency with the secure computation rate requirements by optimizing the CPU cycling frequencies in local execution, the transmit power of IoT equipment, the time-slot allocation and the passive beamforming of IRS. In \cite{xu2022computation}, the use of UAV relay and IRS is considered in order to mitigate the effects of narrow coverage and unfavorable wireless conditions in IoT, gaining the potential in terms of computation enhancement.

Nevertheless, the preceding works \cite{Li21IoTJ, Mao22TVT, xu2022computation} do not fully take advantage of the properties of the IRS for the secure offloading systems, even though the UAV-mounted IRS can enhance the secrecy rate and computing performance. The tractable three-dimensional path planning of UAV-mounted IRS facilitates the multiple degrees of freedom for edge computing performance enhancement. To the best of the authors’ knowledge, the energy-efficient secure offloading system with the aid of the UAV-mounted IRS has not been explored thoroughly, and is still in its early stages, motivating our work presented herein.

\begin{figure}[t!]
\centering
\includegraphics[width=8cm]{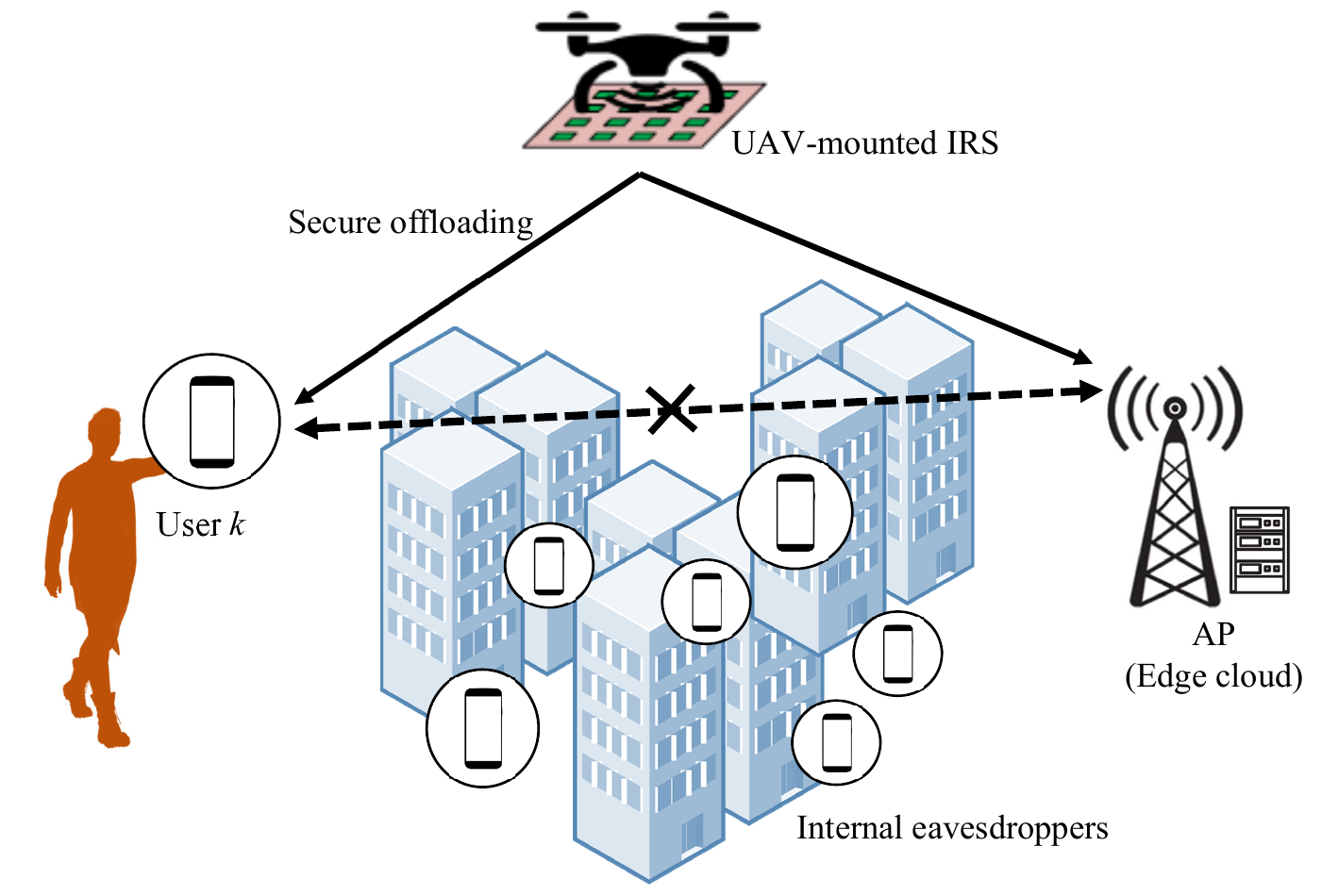}
\caption{System model of energy-efficient secure offloading system designed via UAV-mounted IRS}
\label{fig:fig1}
\end{figure}

In this paper, we develop an energy-efficient secure offloading system, where the UAV-mounted IRS is deployed to create strong virtual links between ground user devices, such as IoT devices, and an access point (AP) acting as an edge cloud. The existing infrastructure may block the LoS links between user devices and the AP, severely hindering data transfer between the input and output interface in complex offloading environments. Furthermore, especially in the IoT scenario, with a number of user devices controlled by the same AP, user privacy and security issues are constantly increasing. Herein, the users, who are not the intended recipients for offloading, are regarded as potential eavesdroppers. We aim to minimize the total energy consumption of battery-limited ground user devices with constraints for secure offloading accomplishment and the operability of the UAV-mounted IRS. To this end, we formulate a joint optimization problem for the transmit power of ground user devices and AP, the trajectory and phase shift matrix of the UAV-mounted IRS and the offloading ratio between local execution and edge computing, whose solutions are obtained based on the successive convex approximation (SCA) method \cite{Scutari14Arxiv}. Via numerical results, it is verified that the proposed algorithm can achieve the considerable energy savings compared to local execution and partial optimizations.

The remaining sections of this paper are outlined as follows. System model and performance metrics are described in Sec. \ref{sec:sys}. In Sec. \ref{sec:opt}, we formulate the optimization problem for the energy-efficient secure offloading system via UAV-mounted IRS, and the proposed algorithm is provided. In addition, we develop the offloading techniques by adopting a more complex aerodynamic model for the flying energy model of the UAV-mounted IRS in Sec. \ref{sec:opt2}. Sec. \ref{sec:num} shows the numerical results with analyses, and the conclusion are provided in Sec. \ref{sec:con}.
\section{System Model}\label{sec:sys}

\subsection{Setup}
We consider a secure offloading system that utilizes an UAV-mounted IRS to assist the offloading procedure from $K$ single-antenna ground users to a single-antenna AP acting as edge server as illustrated in Fig. \ref{fig:fig1}, where the lack of LoS links between the AP and the ground users is assumed. The possible examples and applications can be for instance the communication environments with limited infrastructures in disaster response, emergency relief, military scenarios, or crowded cities with high data traffic. Also, the UAV-mounted IRS can be employed as an alternative option in situations, where the aging infrastructure supporting IoT devices, like terrestrial IRS, has been damaged or destroyed by the extreme climate changes. All the ground users and AP require the secure communication to enable offloading across the same frequency band. However, the use of a cochannel inevitably causes data exposure to undesired recipients. Here, we refer to users, who are not the intended recipients, as the internal eavesdroppers. The UAV-mounted IRS is utilized to strengthen the secrecy of an intended point-to-point connection, as the IRS is supposed to be installed at the UAV, enabling the more flexibility to set up a strong LoS between the nodes than terrestrial IRS \cite{Wu21JSAC}. For the ease of design complexity, we adopt a uniform linear array\footnote{With the minor modifications to the phase shift design, the proposed algorithm in this work can be applied for the uniform planar array of UAV-mounted IRS.} comprising $L$ reflecting elements at the IRS, each of which is controlled by an embedded development board mounted on the UAV.

We assume that a reliable power supply is available at AP, e.g., via plugging to the power supplier or equipping battery storage with sufficient capacity \cite{Che15JSAC}, and the AP does not have any restrictions on the resolution of its digital front-ends. We also suppose that the size of the output bits from the AP is much smaller than the size of the offloading bits required for the operation in each user, which allows us to focus on the energy consumption of uplink transmission utilized in the offloading procedures. For each user $k \in \mathcal{K} =\{1, \dots, K\}$, frequency division duplex (FDD) with an equal division of channel bandwidth is assumed. The input information to be processed is denoted by $I_k$ in bits, and the CPU cycles required per input bit for computation is denoted by $C_k$ in cycles/bits. The mission time is limited to $T$, during which all user requests have to be processed.

For the positions of the ground nodes and the UAV-mounted IRS, we use three-dimensional (3D) Cartesian coordinates in meters (m). The ground nodes, including the users and AP, are deployed at the planar coordinates $\pmb{p}_k=(x_k, y_k)$ and $\pmb{p}_A = (x_A, y_A)$, respectively, in the $xy$-plane, while the UAV-mounted IRS flies along a horizontal trajectory $\pmb{p}_U(t) = (x_U(t), y_U(t))$ with the fixed altitude $H$ for $0 \le t \le T$. The fixed altitude $H$ of UAV-mounted IRS is determined as the altitude that ensures aviation safety, e.g., provision against sudden stall or building collisions \cite{Zhang19TWC}. The UAV-mounted IRS is assumed to be designed compactly so that the same positions are considered for both the UAV and IRS. For tractability, we slice the mission time $T$ into $N_s$ time slots uniformly, such that $T = N_st_s$, where $t_s$ represents the slot length that is set to be small enough so that the ground nodes perceive the position of the UAV-mounted IRS as a constant within each slot. Consequently, the trajectory of the UAV-mounted IRS $\pmb{p}_U(t)$ at each slot can be discretized as $\pmb{p}_U[n]=(x_U[n], y_U[n])$, for $n \in \mathcal{N} =\{1, \dots, N_s\}$, whose horizontal trajectory $(x_U(t), y_U(t))$ over $T$ can be roughly represented by the sequence $\{x_U[n], y_U[n]\}_{n=1}^{N_s}$. Furthermore, owing to the operational capability as well as the starting point and final point of the UAV-mounted IRS, the constraints of the UAV-mounted IRS include its maximum speed as $v_{\max}$ in m/s, and its initial and terminal coordinates as $(x_I, y_I, H)$ and $(x_F, y_F, H)$, where $\pmb{p}_U[1] = (x_I, y_I)$ and $\pmb{p}_U[N_s+1] = (x_F, y_F)$.
Accordingly, the UAV-mounted IRS can fly no further than $D_{\max} = v_{\max}t_s$ within a single time slot. For feasibility, the distance between the initial and terminal coordinates is assumed to satisfy the condition of $\sqrt{(x_F-x_I)^2+(y_F-y_I)^2}\le v_{\max} T$. For each slot, the mobility constraint of UAV-mounted IRS is given as
\begin{equation}\label{eq:Dmin}
||\pmb{p}_U[n+1]-\pmb{p}_U[n]|| \le D_{\max}, \hspace{0.2cm} \text{for}\hspace{0.2cm} n \in \mathcal{N}.
\end{equation}
We define the IRS diagonal phase shift matrix as $\pmb{\Theta}[n] \triangleq \text{diag}\{e^{j\theta_1[n]},$ $e^{j\theta_2[n]}, \dots, e^{j\theta_L[n]}\} \in \mathbb{C}^{L \times L}$ with the $l$th phase shift of the subsurface $\theta_l[n] \in [0, 2\pi]$ for $l \in \{1, \dots, L\}$, where the reflection amplitude of each subsurface is assumed to be one. Here, the finite number $Q$ of discrete levels from $0$ to $2\pi$ is chosen for the phase-shift of subsurface $\theta_l[n]$ to ease the circuit implementation in practice, and accordingly we have $\theta_l \in \mathcal{\eta} \triangleq \{0, \Delta\theta, \dots, (Q-1)\Delta\theta\}$, where $\Delta\theta = 2\pi/Q$.
In the offloading process, we consider either air-to-ground (A2G) or ground-to-air (G2A) channels, where the ground nodes include the AP and $K$ ground users. Following \cite{Li20WCL, Long20Arxiv}, the channel gains from ground node $i$ to the UAV-mounted IRS and from the UAV-mounted IRS to ground node $i$ in the time slot $n$ are expressed as
\begin{subequations}\label{eq:ch}
\begin{eqnarray}
\pmb{h}_{iU}(\pmb{p}_U[n]) \hspace{6.5cm} \nonumber\\
= \sqrt{g_0d_{iU}^{-2}[n]}[1 \,\,\, e^{-jk_wd\phi_{iU}[n]} \,\,\, \cdots \,\,\,e^{-jk_wd(L-1)\phi_{iU}[n]}]^T,\label{eq:G2Ach}\\
\pmb{h}_{Ui}(\pmb{p}_U[n]) \hspace{6.5cm} \nonumber\\
= \sqrt{g_0d_{Ui}^{-2}[n]}[1 \,\,\, e^{-jk_wd\phi_{Ui}[n]} \,\,\, \cdots \,\,\,e^{-jk_wd(L-1)\phi_{Ui}[n]}]^T,\label{eq:A2Gch}
\end{eqnarray}
\end{subequations}
for $i \in \mathcal{K} \cup \{A\}$, respectively, where $g_0$ represents the channel power gain at the reference distance $d_0=1$m, which is determined by the antenna gains of the transceiver and the center frequency, $d$ represents the antenna spacing, and $k_w = 2\pi/\lambda$ is the wavenumber of the carrier defined with the wavelength $\lambda$. In (\ref{eq:ch}), $d_{Ui}[n]=d_{iU}[n]=\sqrt{(x_U[n]-x_i)^2+(y_U[n]-y_i)^2+H^2}$ is the distance between UAV-mounted IRS and ground node $i$ at the time slot $n$, and $\phi_{iU}[n] = (x_U[n]-x_i)/d_{Ui}[n]$ and $\phi_{Ui}[n] = (x_i-x_U[n])/d_{Ui}[n]$ are the cosine of the angle of arrival (AoA) of the signal from ground node $i$ to UAV-mounted IRS and from UAV-mounted IRS to ground node $i$, respectively. We assume an additive white Gaussian noise (AWGN) at the receiver with zero mean and noise power $\sigma^2$. 
For uplink communication, we use $\pmb{h}_{kU}(\pmb{p}_U[n]) \in \mathbb{C}^{L \times 1}$ and $\pmb{h}_{UA}^H(\pmb{p}_U[n]) \in \mathbb{C}^{1 \times L}$ to denote the related baseband channels from user $k$ to UAV-mounted IRS and from UAV-mounted IRS to AP, respectively. Therefore, the effective channel from user $k$ to AP is given by
\begin{equation}\label{eq:effch}
g_{kA}(\pmb{p}_U[n], \pmb{\Theta}[n]) \triangleq \pmb{h}_{UA}^H(\pmb{p}_U[n])\pmb{\Theta}[n]\pmb{h}_{kU}(\pmb{p}_U[n]).
\end{equation}
In addition, we denote the transmit power of the user $k$ for each slot as $\pi_k[n]$, which is supposed to be bounded by the average and peak power constraints over time owing to their operational capability as \vspace{+0.1cm}
\begin{eqnarray} \label{eq:avgmaxpwr}
\frac{1}{N_s}\sum_{n=1}^{N_s}\pi_k[n] \le P_{T}^{\text{avg}}, \quad 0 \le \pi_k[n] \le P_{T}^{\max},
\end{eqnarray}
for $n \in \{1, \dots, N_s\}$.\vspace{+0.1cm}

When there is no internal eavesdropper, the achievable rate from user $k$ to AP at time slot $n$ can be defined as
\vspace{+0.16cm}\begin{eqnarray}
R_{kA}(\pi_k[n], \pmb{p}_U[n], \pmb{\Theta}[n]) \hspace{4cm}&&   \nonumber  \\
= \log_2\left(1+\frac{\pi_k[n]||g_{kA}(\pmb{p}_U[n], \pmb{\Theta}[n])||^2}{\sigma^2}\right).&&\label{eq:Rk2AP}
\end{eqnarray}
If user $j \in \mathcal{K}$ for $j \neq k$ attempts to overhear the uplink communication between user $k$ and AP in time slot $n$, the achievable rate at user $j$ is written as\vspace{+0.16cm}
\begin{eqnarray}
R_{kj}(\pi_k[n], \pmb{p}_U[n], \pmb{\Theta}[n]) \hspace{4cm}&& \nonumber\\
= \log_2\left(1+\frac{\pi_k[n]||g_{kj}(\pmb{p}_U[n], \pmb{\Theta}[n])||^2}{\sigma^2}\right),&&\label{eq:Rk2j}
\end{eqnarray}
 with $g_{kj}(\pmb{p}_U[n], \pmb{\Theta}[n]) \triangleq \pmb{h}_{Uj}^H(\pmb{p}_U[n])\pmb{\Theta}[n]\pmb{h}_{kU}(\pmb{p}_U[n])$. According to \cite{Liang09Eurasip, JSA13WCL}, the secrecy rate between user $k$ and AP at time slot $n$ in uplink can be given as
\begin{subequations}\vspace{+0.15cm}
\begin{eqnarray}
&&\hspace{-1.2cm}R_{kA}^{\text{Up}}(\pi_k[n], \pmb{p}_U[n], \pmb{\Theta}[n]) = \left[R_{kA}(\pi_k[n], \pmb{p}_U[n], \pmb{\Theta}[n]) \right.\hspace{-0.5cm}\nonumber\\
&&\hspace{+0.9cm}- \max_{j \in \mathcal{K}, j \neq k}R_{kj}(\pi_k[n], \pmb{p}_U[n], \pmb{\Theta}[n])]^+, \\
&&\hspace{-0.9cm}= \min_{j \in \mathcal{K}, j \neq k}[R_{kA}(\pi_k[n], \pmb{p}_U[n], \pmb{\Theta}[n]) \nonumber\\
&&\hspace{+0.9cm}- R_{kj}(\pi_k[n], \pmb{p}_U[n], \pmb{\Theta}[n])]^+,\label{eq:upSec}
\end{eqnarray}\vspace{+0.15cm}
\end{subequations}
where $[x]^+=\max\{x, 0\}$.
\subsection{Energy Model for Offloading}
\subsubsection{Communication Energy Model}

In edge computing systems, the communication energy required at $K$ ground users depends on the type of multiple access employed, such as orthogonal or non-orthogonal access (e.g., NOMA in 5G). For system modeling, we focus on orthogonal access, which can be analytically extended to non-orthogonal access by treating undesired interferences as additive noise. In orthogonal access, the uplink communication for each user needs to be performed within $t_s/K$ s, which is equally preallocated. Notably, the energy consumption of the UAV-mounted IRS for communication can be assumed to be zero, since the IRS only reflects passively \cite{Huang19TWC}. Furthermore, compared with transmission energy consumption, the energy consumed during reception is negligible, especially for long distances \cite{Ng11TITS}. Consequently, when the $k$th ground user spends $\pi_k[n]$ at each frame $n$ to offload $I_k$ bits to the AP via UAV-mounted IRS within the mission time $T$, the communication energy consumption at the $n$ slot can be calculated as 
\begin{equation}\label{eq:commE}
E_k^{\text{Comm}}(\pi_k[n])=\frac{\pi_k[n]t_s}{K},
\end{equation}
for $k \in \mathcal{K}$.

\subsubsection{Flying Energy Model}\label{sec:flyingE}
Similar to \cite{JSA18TVT, Borst08Air}, the flying energy model for UAV-mounted IRS at each frame $n$ can be written as
\begin{equation}\label{eq:flyingE}
E_U^{\text{F}}(\pmb{v}_U[n])=\frac{1}{2} m_u t_s ||\pmb{v}_U[n]||^2,
\end{equation}
where $m_u$ is the total mass of the UAV-mounted IRS that includes its payload, and $\pmb{v}_U[n]$ represents the velocity of the UAV-mounted IRS, and is defined as
\begin{equation}\label{eq:velocity}
\pmb{v}_U[n]=\frac{\sqrt{(x_U[n+1]-x_U[n])^2+(y_U[n+1]-y_U[n])^2}}{t_s}.
\end{equation}
Here, the flying energy model in (\ref{eq:flyingE}) relies entirely on the velocity vector $\pmb{v}_U[n]$, which only accounts for the kinetic energy as the UAV-mounted IRS is assumed to be in level flight with no gravitational potential energy.
\vspace{-0.2cm}
\subsection{Energy Model for Local Execution}
Contrary to the constrained ground user devices, the AP is supposed to have a sufficient energy budget as an edge server. For reference, we investigate the energy consumption of the local execution case, where all tasks are locally computed at user devices solely. To process $I_k$ bits at each user $k$ within $T$ s, the CPU frequency $f_k$ of the $k$th user has to be set to $C_kI_k/T$. The total energy consumption of local execution is then given by
\begin{equation}\label{eq:localE}
E^\text{Local}_k(I_k)\triangleq \gamma_k C_k I_k (f_k)^2 = \gamma_k \frac{C_k^3 I_k^3}{T^2},
\end{equation} 
,where $\gamma_k$ represents effective switched capacitance of user $k$, which is derived as in \cite{Yuan03ACM, Yuan06ACM}.
\section{Optimal Energy Consumption of Secure Offloading System}\label{sec:opt}
In this paper, our objective is to minimize the total energy consumption of the battery-limited $K$ ground user devices by jointly optimizing their transmit powers, trajectory and phase shift matrix of UAV-mounted IRS and the offloading ratio between local execution and edge computing under the constraints for uplink secrecy rate as well as for the energy budget and mobility of UAV-mounted IRS. For this purpose, the optimization problem can be formulated as
\begin{subequations}\label{eq:opt1}
\begin{eqnarray}
&&\hspace{-1.5cm} \min_{\pmb{\pi}, \pmb{p}_U, \pmb{v}_U,\pmb{\Theta},\pmb{\rho}} \hspace{0.2cm} \sum_{k=1}^K\sum_{n=1}^{N_s}E_k^{\text{Comm}}(\pi_k[n]) + \sum_{k=1}^K E^\text{Local}_k(\rho_k I_k) \label{eq:obj}\\
&&\hspace{-1.5cm} \text{s.t.} \hspace{0.2cm} \sum_{n=1}^{N_s}B_{kA}R_{kA}^{\text{Up}}(\pi_k[n], \pmb{p}_U[n], \pmb{\Theta}[n]) \ge I_k (1-\rho_k), \hspace{0.2cm} \nonumber\\
&&\hspace{+4.0cm} \text{for} \hspace{0.2cm} k \in \mathcal{K}, \label{eq:upConst}\\
&&\hspace{-0.8cm}  \sum_{n=1}^{N_s}E_{U}^{\text{F}}(\pmb{v}_U[n]) \le E_{th},\hspace{0.2cm} \text{for} \hspace{0.2cm} n \in \mathcal{N}, \label{eq:UAVConst}\\
&&\hspace{-0.8cm}  \pmb{v}_U[n]=\frac{||\pmb{p}_U[n+1]-\pmb{p}_U[n]||}{t_s}, \hspace{0.2cm} \text{for} \hspace{0.2cm} n \in \mathcal{N}, \label{eq:vconst}\\
&&\hspace{-0.8cm} \pmb{p}_U[1] = (x_I, y_I), \hspace{0.2cm} \pmb{p}_U[N_s+1] = (x_F, y_F),\label{eq:posConst}\\
&&\hspace{-0.8cm}  \pi_k[n] \ge 0, \hspace{0.2cm} \text{for} \hspace{0.2cm} k \in \mathcal{K} \hspace{0.2cm} \text{and} \hspace{0.2cm} n \in \mathcal{N}, \label{eq:pwrconst}\\
&&\hspace{-0.8cm}  0 \le \rho_k \le 1, \hspace{0.2cm} \text{for} \hspace{0.2cm} k \in \mathcal{K}, \label{eq:rhoconst}\\
&&\hspace{-0.8cm}  \text{(\ref{eq:Dmin}) \text{and} \text{(\ref{eq:avgmaxpwr})}}, \label{eq:opt1remaining}
\end{eqnarray}
\end{subequations}
where $\pmb{\pi} = \{\pi_k[n]\}_{\forall k, n},\ \pmb{p}_U = \{\pmb{p}_U[n]\}_{\forall n},\ \pmb{v}_U = \{\pmb{v}_U[n]\}_{\forall n}, $ $\pmb{\Theta} = \{\pmb{\Theta}[n]\}_{\forall n}$ and $\pmb{\rho}=\{\rho_k\}_{\forall k}$, with $0 \le \rho_k \le 1$ being the offloading ratio between local execution and offloading for the ground user $k$. For example, $\rho_k=1$ indicates the complete local execution to compute the task entirely at ground user $k$, while $\rho_k=0$ indicates the complete offloading to execute the task entirely at AP. In (\ref{eq:opt1}), $B_{kA}$ is the bandwidth allocated for the uplink communication between the $k$th ground user and AP, and $E_{th}$ represents the energy budget of the UAV-mounted IRS within the given mission time. 
The optimization problem (\ref{eq:opt1}) is nonconvex since the uplink secrecy rate constraint (\ref{eq:upConst}) involves coupled term with power control variables $\pmb{\pi}$, trajectory variables $\pmb{p}_U$ and phase shift matrix $\pmb{\Theta}$. To address the constraint (\ref{eq:upConst}), we first introduce the slack variable $\pmb{s}=\{s_k\}_{\forall k}$, and reformulate (\ref{eq:opt1}) as
\begin{subequations}\label{eq:opt2}
\begin{eqnarray}
&&\hspace{-1.5cm} \min_{\pmb{\pi}, \pmb{p}_U, \pmb{v}_U, \pmb{\Theta}, \pmb{\rho}, \pmb{s} \hspace{0.2cm}}\sum_{k=1}^K\sum_{n=1}^{N_s}E_k^{\text{Comm}}(\pi_k[n]) + \sum_{k=1}^K E^\text{Local}_k(\rho_kI_k)\label{eq:obj2}\\
&&\hspace{-1.5cm} \text{s.t.} \hspace{0.2cm} s_k \ge I_k(1-\rho_k), \hspace{0.2cm} \text{for} \hspace{0.2cm} k \in \mathcal{K}, \label{eq:s2}\\
&&\hspace{-0.9cm}  \sum_{n=1}^{N_s}  B_{kA}\left[R_{kA}(\pi_k[n], \pmb{p}_U[n], \pmb{\Theta}[n]) \right.\nonumber\\
&&\hspace{+1.5cm} \left. - R_{kj}(\pi_k[n], \pmb{p}_U[n], \pmb{\Theta}[n])\right]^+ \ge s_k, \nonumber \\
&&\hspace{+3.0cm} \text{for} \hspace{0.2cm} k,j \in \mathcal{K}\ \text{and}\ j \neq k, \label{eq:upConst2}\\
&&\hspace{-0.9cm} \text{(\ref{eq:UAVConst}) - (\ref{eq:opt1remaining})}.
\end{eqnarray}
\end{subequations}
To handle the non-convexity of (\ref{eq:opt2}) due to coupling, we propose an alternative algorithm based on the SCA approach \cite{Scutari14Arxiv, Beck10Book} to provide the suboptimal solution of (\ref{eq:opt2}). In the following, the subproblems required for the proposed alternating method are provided.
\subsection{Joint Phase and Trajectory Optimization}\label{sec:phaseTOpt}
To make the problem (\ref{eq:opt2}) tractable, we fix the transmit power $\pmb{\pi}$ of the ground users, offloading ratio $\pmb{\rho}$ and slack variable $\pmb{s}$ in (\ref{eq:opt2}), which are regarded as deterministic values in the following. For any given $\pmb{\pi}$, $\pmb{\rho}$, and $\pmb{s}$, we can rewrite the problem (\ref{eq:opt2}) as
\begin{subequations}\label{eq:phaseTopt}
\begin{eqnarray}
&&\hspace{-4.0cm} \text{Find} \hspace{0.2cm} \pmb{p}_U, \pmb{\Theta} \label{eq:phaseTobj}\\
&&\hspace{-4.0cm} \text{s.t.} \hspace{0.2cm} \text{(\ref{eq:upConst2}), (\ref{eq:UAVConst})-(\ref{eq:posConst}), (\ref{eq:Dmin})}.
\end{eqnarray}
\end{subequations}
It is noted in (\ref{eq:phaseTopt}) that the objective function of (\ref{eq:phaseTopt}) does not depend on the trajectory $\pmb{p}_U$ and phase matrix $\pmb{\Theta}$ of UAV-mounted IRS. Moreover, the problem (\ref{eq:phaseTopt}) is still nonconvex owing to the nonconvex constraints (\ref{eq:upConst2}). To address this issue, we use a two-step approach, where the trajectory is obtained after optimizing the phase shift matrix.
\subsubsection{Phase Optimization}
The phase optimization can be implemented with the several bounds and approximations by using the convex relaxation techniques, which can degrade the optimality performance \cite{Long20Arxiv}. Therefore, we adopt a simple but powerful phase design to adjust the phases of the received signal at the AP in the uplink to maximize the received signal strength for the secrecy rate constraint (\ref{eq:upConst2}). For any given trajectory $\pmb{p}_U$, the effective channel $g_{kA}(\pmb{p}_U[n], \pmb{\Theta}[n])$ for uplink can be expressed as
\begin{eqnarray}\label{eq:effch_user}
&&\hspace{-1cm}g_{kA}(\pmb{p}_U[n], \pmb{\Theta}[n]) \nonumber\\
&&\hspace{-0.7cm}= \frac{g_0\sum_{l=1}^Le^{j \left(\theta_l[n] + k_wd(l-1)\left(\phi_{UA}[n]-\phi_{kU}[n]\right)\right)}}{\sqrt{d_{UA}^2[n]d_{kU}^2[n]}}.
\end{eqnarray}
To maximize the received signal strength, we need to coherently combine the signals of the different paths at the AP in the uplink, which yields
\begin{eqnarray}\label{eq:optphase}
\theta_{l}[n] = k_wd(l-1)\left(\phi_{kU}[n]-\phi_{UA}[n]\right) + \omega,
\end{eqnarray}
$\forall l, n, k$, respectively, where $\omega \in [0, 2\pi]$. By using (\ref{eq:effch_user}) and (\ref{eq:optphase}), we can obtain the effective channel of uplink as
\begin{equation}\label{eq:effch_user2}
g_{kA}(\pmb{p}_U[n]) = \frac{g_0Le^{j \omega}}{\sqrt{d_{UA}^2[n]d_{kU}^2[n]}},
\end{equation}
which depends only on trajectory $\pmb{p}_U[n]$, and yields the achievable rate $R_{kA}(\pi_k[n], \pmb{p}_U[n])$ in (\ref{eq:upSec}) at AP for uplink as
\begin{equation}\label{eq:trajrate_user}
\check{R}_{kA}(\pi_k[n], \pmb{p}_U[n]) \triangleq \log_2\left(1+\frac{\pi_k[n]g_0^2L^2}{\sigma^2d_{UA}^2[n]d_{kU}^2[n]}\right).
\end{equation}

With the closed-form phase shifts in (\ref{eq:optphase}), the effective channel $g_{kj}(\pmb{p}_U[n], \pmb{\Theta}[n])$ at the internal eavesdropper $j$ for $j \neq k$ is similarly written as (c.f. (\ref{eq:effch_user})-(\ref{eq:effch_user2}))
\begin{eqnarray}\label{eq:effch_eve}
&&\hspace{-1cm}\|g_{kj}(\pmb{p}_U[n], \pmb{\Theta}[n])\|^2 \nonumber \\
&&\hspace{-0.7cm}= \|\frac{g_0\sum_{l=1}^Le^{j \left(\theta_l[n] + k_wd(l-1)\left(\phi_{Uj}[n]-\phi_{kU}[n]\right)\right)}}{\sqrt{d_{Uj}^2[n]d_{kU}^2[n]}}\|^2 \nonumber \\
&&\hspace{-0.7cm}\le \frac{g_0^2L^2}{d_{Uj}^2[n]d_{kU}^2[n]},
\end{eqnarray}
where the last upper bound can be obtained when the different paths at the internal eavesdropper $j$ are combined coherently. Consequently, the achievable rate $R_{kj}(\pi_k[n], \pmb{p}_U[n], \pmb{\Theta}[n])$ in uplink can be upper-bounded as
\begin{eqnarray}\label{eq:trajrate_eve}
&& \hspace{-1.4cm} R_{kj}(\pi_k[n], \pmb{p}_U[n], \pmb{\Theta}[n]) \nonumber\\
&& \hspace{-1.1cm}\le \log_2\left(1+\frac{g_0^2L^2\pi_k[n]}{\sigma^2d_{Uj}^2[n]d_{kU}^2[n]}\right) \triangleq \check{R}_{kj}(\pi_k[n], \pmb{p}_U[n]).
\end{eqnarray}
\subsubsection{Trajectory Optimization}
Based on the results (\ref{eq:trajrate_user}) and (\ref{eq:trajrate_eve}), we design the trajectory $\pmb{p}_U$ and the velocity $\pmb{v}_U$ of UAV-mounted IRS in this section. To handle the nonconvexity of constraint (\ref{eq:upConst2}) with respect to $\pmb{p}_U$, we introduce the slack variables $\pmb{q} =\{q_k[n]\}_{\forall k, n}$ and $\pmb{r}= \{r_{kj}[n]\}_{\forall k, j \neq k, n}$. Then, the problem (\ref{eq:phaseTopt}) can be represented
\begin{subequations}\label{eq:phaseTopt2}
\begin{eqnarray}
&&\hspace{-1.2cm} \text{Find} \hspace{0.2cm} \pmb{p}_U, \pmb{v}_U,\pmb{q}, \pmb{r} \label{eq:phaseTobj2}\\
&&\hspace{-1.2cm} \text{s.t.} \hspace{0cm} \sum_{n=1}^{N_s}B_{kA}\left[\check{R}_{kA}(\pi_k[n], q_k[n]) - \check{R}_{kj}(\pi_k[n], r_{kj}[n])\right]^+ \nonumber\\
&&\hspace{2.35cm} \ge s_k, \hspace{0.2cm} \text{for} \hspace{0.2cm} k,j \in \mathcal{K}, j \neq k, \label{eq:upConst2re} \\
&&\hspace{-0.7cm} d_{UA}^2[n]d_{kU}^2[n] \le q_k[n],  \hspace{0.2cm} \text{for} \hspace{0.2cm} k \in \mathcal{K} \hspace{0.2cm} \text{and} \hspace{0.2cm} n \in \mathcal{N}, \label{eq:qk}\\
&&\hspace{-0.7cm} d_{Uj}^2[n]d_{kU}^2[n] \ge r_{kj}[n],  \hspace{0.2cm} \nonumber\\
&&\hspace{1.6cm} \text{for} \hspace{0.2cm} k, j \in \mathcal{K}, j \neq k \hspace{0.2cm} \text{and} \hspace{0.2cm} n \in \mathcal{N}.  \label{eq:rkj}\\
&&\hspace{-0.7cm}  \text{(\ref{eq:UAVConst})-(\ref{eq:posConst}), (\ref{eq:Dmin})},
\end{eqnarray}
\end{subequations}
where we have defined
\begin{subequations}
\begin{eqnarray}
&&\hspace{-1.2cm}\check{R}_{kA}(\pi_k[n], q_k[n])\nonumber\\ 
&&\hspace{-0.9cm}=\log_2\left(\sigma^2q_k[n]+g_0^2L^2\pi_k[n]\right)-\log_2\left(\sigma^2q_k[n]\right),\\
&&\hspace{-1.2cm}\check{R}_{kj}(\pi_k[n], r_{kj}[n])\nonumber\\
&&\hspace{-0.9cm}= \log_2\left(\sigma^2r_{kj}[n]+g_0^2L^2\pi_k[n]\right)-\log_2\left(\sigma^2r_{kj}[n]\right).
\end{eqnarray}
\end{subequations}
The problem (\ref{eq:phaseTopt2}) is nonconvex because the constraints (\ref{eq:upConst2re}) - (\ref{eq:rkj}) have a nonconvex structure, with the difference of concave (DC) structure in (\ref{eq:upConst2re}) and the coupling constraints in (\ref{eq:qk}) - (\ref{eq:rkj}). By using the SCA approach, we successively solve the series of convex optimization problems obtained from (\ref{eq:phaseTopt2}). 

Specifically, we first define the set of primal variables for the problem (\ref{eq:phaseTopt2}) as $\pmb{\Gamma} = \{\pmb{\Gamma}[n]\}_{\forall n}$ with 
$\pmb{\Gamma}[n]=(\pmb{p}_U[n], \{q_k[n]\}_{\forall k}$, $\{r_{kj}[n]\}_{\forall k, j \neq k, n})$, and define the 
$\pmb{\Gamma}^{(z)}=\{\pmb{\Gamma}^{(z)}[n]\}_{\forall n}$
 with $\pmb{\Gamma}^{(z)}[n]=(\pmb{p}_U^{(z)}[n], \{q_k^{(z)}[n]\}_{\forall k}, \{r_{kj}^{(z)}[n]\}_{\forall k, j \neq k, n})\in \pmb{\Omega}$ for the $z$th iterate within the feasible set $\pmb{\Omega}$ of (\ref{eq:phaseTopt2}). Then, to simplify the notation, we define $\pmb{p}_U^{(z)}=\{\pmb{p}_U^{(z)}[n]\}_{\forall n},\ \pmb{q}^{(z)}=\{q_k^{(z)}[n]\}$ $_{\forall k, n}$ and $\pmb{r}^{(z)}=\{r_{kj}^{(z)}[n]\}_{\forall k, j \neq k, n}$. To tackle the nonconvex constraint with the DC structure, we use the following lemma.
\begin{lemma}\label{lem1}
(\cite[Example 3]{Scutari14Arxiv}) For the nonconvex constraint of the form $f(\pmb{x}) = f^+(\pmb{x}) - f^-(\pmb{x}) \ge 0$, where functions $f^+(\pmb{x})$ and $f^-(\pmb{x})$ are both concave and continuously differentiable, we can use the convex lower approximation of $f(\pmb{x})$ under the conditions \cite[Assumption 3]{Scutari14Arxiv}, which is essential for the SCA technique. Then, we have 
\begin{eqnarray}\label{eq:dc}
\bar{f}(\pmb{x}; \pmb{\mu}) \triangleq f^+(\pmb{x}) - f^-(\pmb{\mu}) - \nabla_{\pmb{x}}f^-(\pmb{\mu})^T(\pmb{x}-\pmb{\mu}) \le f(\pmb{x}).
\end{eqnarray}
The lower bound (\ref{eq:dc}) is obtained by linearizing the convex term $- f^-(\pmb{x})$ and leaving the concave term $f^+(\pmb{x})$ for all $\pmb{x}$ in the domain of $f(\pmb{x})$ and $\pmb{\mu}$ in the feasible set.
\end{lemma}

By using Lemma \ref{lem1}, the DC function in (\ref{eq:upConst2re}) is lower-bounded as
\begin{eqnarray}
&&\hspace{-1.2cm}\check{R}_{kA}(\pi_k[n], q_k[n]) -\check{R}_{kj}(\pi_k[n], r_{kj}[n]) \nonumber\\
&&\hspace{-0.9cm}\ge \log_2\left(\sigma^2q_k[n]+g_0^2L^2\pi_k[n]\right)+ \log_2\left(\sigma^2r_{kj}[n]\right)\nonumber\\
&&\hspace{-0.6cm}-\log_2\left(\sigma^2q_k^{(z)}[n]\right)- \frac{\left(q_k[n]-q_k^{(z)}[n]\right)}{\ln 2 q_k^{(z)}[n]}\nonumber\\
&&\hspace{-0.6cm}  - \log_2\left(\sigma^2r_{kj}^{(z)}[n]+g_0^2L^2\pi_k[n]\right)\nonumber\\
&&\hspace{-0.6cm}- \frac{\sigma^2\left(r_{kj}[n]-r_{kj}^{(z)}[n]\right)}{\ln 2 \left(\sigma^2r_{kj}^{(z)}[n]+g_0^2L^2\pi_k[n]\right)}\nonumber\\
&&\hspace{-0.9cm}\triangleq \check{R}_{kA,j}^{\text{Up}}(\pmb{\Gamma}[n], \pmb{\Gamma}^{(z)}[n]). \label{eq:upConst4}
\end{eqnarray}

Moreover, the nonconvex constraints (\ref{eq:qk}) and (\ref{eq:rkj}) are the multiplication of convex and nonnegative functions, which can be handled based on the following Lemma and Remark.
\begin{lemma} \label{lem2}
(\cite[Example 4]{Scutari14Arxiv}) To address the nonconvexity of the problem with a nonconvex constraint in bi-convex structure, i.e., $h(\pmb{x}_1, \pmb{x}_2) = p_1(\pmb{x}_1)p_2(\pmb{x}_2) \le 0$, where $p_1(\pmb{x}_1)$ and $p_2(\pmb{x}_2)$ are both convex and nonnegative for all $(\pmb{x}_1, \pmb{x}_2)$ in the domain of $h(\pmb{x}_1, \pmb{x}_2)$ and $(\pmb{\mu}_1, \pmb{\mu}_2)$ in the feasible set, the following upper approximation of $h(\pmb{x}_1, \pmb{x}_2)$ under the conditions \cite[Assumption 3]{Scutari14Arxiv} is given as
\begin{eqnarray}
&&\hspace{-1.2cm} \bar{h}(\pmb{x}_1, \pmb{x}_2; \pmb{\mu}_1, \pmb{\mu}_2) \nonumber\\
&&\hspace{-0.9cm} \triangleq \frac{1}{2}\left(p_1(\pmb{x}_1) + p_2(\pmb{x}_2)\right)^2 - \frac{1}{2}\left(p^2_1(\pmb{\mu}_1) + p^2_2(\pmb{\mu}_2)\right)\nonumber\\
&&\hspace{-0.6cm} - p_1(\pmb{\mu}_1)p_1'(\pmb{\mu}_1)(\pmb{x}_1-\pmb{\mu}_1)-p_2(\pmb{\mu}_2)p_2'(\pmb{\mu}_2)(\pmb{x}_2-\pmb{\mu}_2)\nonumber\\
&&\hspace{-0.9cm} \ge h(\pmb{x}_1, \pmb{x}_2).
\end{eqnarray}    
\end{lemma}

\begin{remark}\label{rem1}
Based on Lemma \ref{lem2}, the nonconvex constraint $h(\pmb{x}_1, \pmb{x}_2) = p_1(\pmb{x}_1)p_2(\pmb{x}_2) \ge 0$, where $h(\pmb{x}_1, \pmb{x}_2)$ is the coupled term by convex and nonnegative functions, $p_1(\pmb{x}_1)$ and $p_2(\pmb{x}_2)$, for all in the domain of $h(\pmb{x}_1, \pmb{x}_2)$ and $(\pmb{\mu}_1, \pmb{\mu}_2)$ in the feasible set, can be handled with the following convex approximation:
\begin{eqnarray}
&&\hspace{-1.5cm} \bar{h}(\pmb{x}_1, \pmb{x}_2; \pmb{\mu}_1, \pmb{\mu}_2)\nonumber\\
&&\hspace{-1.2cm}\triangleq \frac{1}{2}\left(p_1(\pmb{\mu}_1) + p_2(\pmb{\mu}_2)\right)^2 - \frac{1}{2}\left(p^2_1(\pmb{x}_1) + p^2_2(\pmb{x}_2)\right)\nonumber\nonumber\\
&&\hspace{-0.9cm} +\left(p_1(\pmb{\mu}_1)+p_2(\pmb{\mu}_2)\right)\left(p_1'(\pmb{\mu}_1)(\pmb{x}_1-\pmb{\mu}_1)\right.\nonumber\\
&&\hspace{-0.9cm}\left.+p_2'(\pmb{\mu}_2)(\pmb{x}_2-\pmb{\mu}_2)\right)\le h(\pmb{x}_1, \pmb{x}_2).
\end{eqnarray}
\end{remark}
Using Lemma \ref{lem2} and Remark \ref{rem1}, the nonconvex constraints (\ref{eq:qk}) and (\ref{eq:rkj}) can be bounded as
\begin{eqnarray}\label{eq:trajdc2}
&&\hspace{-1.3cm} d_{UA}^2[n]d_{kU}^2[n] \le \frac{1}{2}\left(d_{UA}^2[n] + d_{kU}^2[n]\right)^2 \nonumber\\
&&\hspace{-0.7cm} - \frac{1}{2}\left((d_{UA}^{(z)}[n])^4 + (d_{kU}^{(z)}[n])^4\right)\nonumber\\
&&\hspace{-0.7cm} - 2(d_{UA}^{(z)}[n])^2\left(\pmb{p}_U^{(z)}[n]-\pmb{p}_A\right)^T\left(\pmb{p}_U[n]-\pmb{p}_U^{(z)}[n]\right)\nonumber\\
&&\hspace{-0.7cm} - 2(d_{kU}^{(z)}[n])^2\left(\pmb{p}_U^{(z)}[n]-\pmb{p}_k\right)^T\left(\pmb{p}_U[n]-\pmb{p}_U^{(z)}[n]\right)\nonumber\\
&&\hspace{-1.0cm} \triangleq f_q({\pmb{\Gamma}[n], \pmb{\Gamma}^{(z)}[n]}),\label{eq:qk2}
\end{eqnarray}
\begin{eqnarray}
&&\hspace{-0.6cm} d_{Uj}^2[n]d_{kU}^2[n] \ge \frac{1}{2}\left((d_{Uj}^{(z)}[n])^2 + (d_{kU}^{(z)}[n])^2\right)^2 \nonumber\\
&&\hspace{-0.0cm}- \frac{1}{2}\left(d_{Uj}^4[n] + d_{kU}^4[n]\right)\nonumber\\
&&\hspace{-0.0cm}+ 2(d_{Uj}^{(z)}[n])^2\left(2\pmb{p}_U^{(z)}[n]-\pmb{p}_j-\pmb{p}_k\right)^T\left(\pmb{p}_U[n]-\pmb{p}_U^{(z)}[n]\right)\nonumber\\
&&\hspace{-0.0cm}+ 2(d_{kU}^{(z)}[n])^2\left(2\pmb{p}_U^{(z)}[n]-\pmb{p}_j-\pmb{p}_k\right)^T\left(\pmb{p}_U[n]-\pmb{p}_U^{(z)}[n]\right)\nonumber\\
&&\hspace{-0.3cm} \triangleq f_r({\pmb{\Gamma}[n], \pmb{\Gamma}^{(z)}[n]}),\label{eq:rkj2}
\end{eqnarray}
\begin{algorithm} [t!]
\begin{algorithmic}
\caption{Joint phase and trajectory optimization of UAV-mounted IRS} \label{al1}
\State {\textbf{Input}}: $\pmb{\Gamma}^{(0)} = (\pmb{p}^{(0)}_U, \pmb{q}^{(0)}, \pmb{r}^{(0)})\in \pmb{\Omega}$ and the fixed $\pmb{\pi}, \pmb{s}, \pmb{\rho}$. Set $z=0$.
\State {\textbf{Repeat} until a convergence criterion is satisfied}
\State \hspace{+0.2cm} With given $\pmb{p}_U^{(z)}$, obtain $\pmb{\Theta}^{*}$ using (\ref{eq:optphase}).
\State \hspace{+0.2cm} Find $(\pmb{\Gamma}^*(\pmb{\Gamma}^{(z)}), \pmb{v}_U^*)$ using (\ref{eq:phaseTopt3}).
\State \hspace{+0.2cm} Set $\pmb{\Gamma}^{(z+1)}=\pmb{\Gamma}^{(z)}+\alpha^{(z)}(\pmb{\Gamma}^*(\pmb{\Gamma}^{(z)})-\pmb{\Gamma}^{(z)})$,\hspace{+0.1cm} $\alpha^{(z)} \in (0, 1]$.
\State \hspace{+0.2cm} Update $z \gets z+1$
\State {\textbf{Output}}: $\pmb{\Gamma}^*(\pmb{\Gamma}^{(z)}) = \pmb{p}_U^*, \pmb{v}_U^*, \pmb{\Theta}^{*}$.
\end{algorithmic}
\end{algorithm}respectively. With (\ref{eq:upConst4}), (\ref{eq:qk2}) and (\ref{eq:rkj2}), the problem (\ref{eq:phaseTopt2}) can be finally reformulated as
\begin{subequations}\label{eq:phaseTopt3}
\begin{eqnarray}
&&\hspace{-1.5cm} \text{Find} \hspace{0.2cm} \pmb{\Gamma}, \pmb{v}_U \label{eq:phaseTobj3}\\
&&\hspace{-1.5cm} \text{s.t.} \hspace{0.2cm} \sum_{n=1}^{N_s}B_{kA}\left[\check{R}_{kA,j}^{\text{Up}}(\pmb{\Gamma}[n], \pmb{\Gamma}^{(z)}[n])\right]^+ \ge s_k, \nonumber\\
&&\hspace{1.4cm} \text{for} \hspace{0.2cm} k, j \in \mathcal{K}, j \neq k \hspace{0.2cm} \text{and} \hspace{0.2cm} n \in \mathcal{N},  \label{eq:upConst3re}\\
&&\hspace{-1.2cm} f_q({\pmb{\Gamma}[n], \pmb{\Gamma}^{(z)}[n]}) \le q_k[n], \hspace{0.2cm} \text{for} \hspace{0.2cm} k \in \mathcal{K} \hspace{0.2cm} \text{and} \hspace{0.2cm} n \in \mathcal{N}, \label{eq:qk3}\\
&&\hspace{-1.2cm} f_r({\pmb{\Gamma}[n], \pmb{\Gamma}^{(z)}[n]}) \ge r_{kj}[n], \hspace{0.2cm} \nonumber\\
&&\hspace{1.4cm} \text{for} \hspace{0.2cm} k, j \in \mathcal{K}, j \neq k \hspace{0.2cm} \text{and} \hspace{0.2cm} n \in \mathcal{N}, \label{eq:rkj3}\\
&&\hspace{-1.2cm}  \text{(\ref{eq:UAVConst})-(\ref{eq:posConst}), (\ref{eq:Dmin})}.
\end{eqnarray}
\end{subequations}
Here, by transforming all the constraints in (\ref{eq:phaseTopt3}) as convex, the problem (\ref{eq:phaseTopt3}) can be numerically solved through the standard convex optimization strategies or toolboxes like CVX MOSEK. As the objective function of (\ref{eq:phaseTopt3}) is constant, the solution of problem (\ref{eq:phaseTopt3}) can be randomly or arbitrarily chosen within the feasible set of (\ref{eq:phaseTopt3}). Therefore, the trajectory of UAV-mounted IRS can be obtained using Algorithm \ref{al1}, where the phase of UAV-mounted IRS is designed based on (\ref{eq:optphase}).
\subsection{Joint Power and Offloading Ratio Optimization}\label{sec:powOpt}
In this section, we optimize the power allocation and offloading ratio of the user by considering the given trajectory $\pmb{p}_U$, velocity $\pmb{v}_U$ and phase shift matrix $\pmb{\Theta}$ as deterministic values. The joint power and offloading ratio optimization problem can then be reformulated as 
\begin{subequations}\label{eq:pwropt}
\begin{eqnarray}
&&\hspace{-2cm} \min_{\pmb{\pi}, \pmb{\rho}, \pmb{s}} \hspace{0.2cm} \sum_{k=1}^K\sum_{n=1}^{N_s} \frac{\pi_k[n]t_s}{K} + \sum_{k=1}^K \gamma_k \frac{C_k^3 (\rho_kI_k)^3}{T^2}\label{eq:pwrobj}\\
&&\hspace{-2cm} \text{s.t.} \hspace{0.2cm} \text{(\ref{eq:s2}), (\ref{eq:upConst2})}, \text{(\ref{eq:pwrconst}), (\ref{eq:rhoconst}), (\ref{eq:avgmaxpwr})}.
\end{eqnarray}
\end{subequations}
The problem (\ref{eq:pwropt}) is nonconvex as the left-hand side of the uplink secrecy rate constraint (\ref{eq:upConst2}) has a DC structure with respect to power variable $\pmb{\pi}$. To solve (\ref{eq:pwropt}), we adopt the SCA method as described in Sec. \ref{sec:phaseTOpt}. Then, it is essential to define the set of primal variables for the problem (\ref{eq:pwropt}) as $\pmb{\Gamma} = \{\pmb{\Gamma}[n]\}_{\forall n}$ with $\pmb{\Gamma}[n] = (\{\pi_k[n]\}_{\forall k})$, and define $\pmb{\Gamma}^{(z)} = \{\pmb{\Gamma}^{(z)}[n]\}_{\forall n}$ with $\pmb{\Gamma}^{(z)}[n] = (\{\pi_k^{(z)}[n]\}_{\forall k})\in \pmb{\Omega}$ for the $z$th iterate within the feasible set $\pmb{\Omega}$ of (\ref{eq:pwropt}). Also, 
$\pmb{\pi}^{(z)}$ $=\{\pi_k^{(z)}[n]\}_{\forall k, n}$ is defined to ease the notation. By following Lemma \ref{lem1}, we can regard $R_{kj}(\pi_k[n], \pmb{p}_U[n], \pmb{\Theta}[n])$ in (\ref{eq:upConst2}) as the convex term $- f^-(\pmb{x})$ with respect to $\pi_k[n]$, for the fixed $\pmb{p}_U$ and $\pmb{\Theta}$, which is upper-bounded as 
\begin{eqnarray}
&&\hspace{-1.2cm}R_{kj}(\pmb{\Gamma}[n])\triangleq R_{kj}(\pi_k[n], \pmb{p}_U[n], \pmb{\Theta}[n])\hspace{2.5cm}\nonumber\\
&&\hspace{+0.4cm}= \log_2\left(1+\frac{\pi_k[n]||g_{kj}(\pmb{p}_U[n], \pmb{\Theta}[n])||^2}{\sigma^2}\right) \hspace{0.5cm} \nonumber\\
&&\hspace{+0.4cm}\le \log_2\left(1+\frac{\pi_k^{(z)}[n]||g_{kj}(\pmb{p}_U[n], \pmb{\Theta}[n])||^2}{\sigma^2}\right) \hspace{0.2cm}\nonumber\\
&&\hspace{+0.7cm}+ \frac{||g_{kj}(\pmb{p}_U[n], \pmb{\Theta}[n])||^2\left(\pi_k[n] - \pi_k^{(z)}[n]\right)}{\ln 2\left(\sigma^2 + \pi_k^{(z)}[n]||g_{kj}(\pmb{p}_U[n], \pmb{\Theta}[n])||^2\right)} \hspace{0.1cm} \nonumber\\
&&\hspace{+0.4cm}= \hat{R}_{kj}(\pi_k[n], \pmb{p}_U[n], \pmb{\Theta}[n], \pi_k^{(z)}[n]) \nonumber\\
&&\hspace{+0.4cm}\triangleq \hat{R}_{kj}(\pmb{\Gamma}[n]; \pmb{\Gamma}^{(z)}[n]). \hspace{1.8cm}\label{eq:MMup}
\end{eqnarray}
With the upper bound in (\ref{eq:MMup}), we can reformulate the problem (\ref{eq:pwropt}) as
\begin{subequations}\label{eq:pwropt2}
\begin{eqnarray}
&& \hspace{-0.8cm}\min_{\pmb{\Gamma}} \hspace{0.2cm} \sum_{k=1}^K\sum_{n=1}^{N_s} \frac{\pi_k[n]t_s}{K} + \sum_{k=1}^K \gamma_k \frac{C_k^3 (\rho_kI_k)^3}{T^2}\label{eq:pwrobj2}\\
&& \hspace{-0.8cm}\text{s.t.} \hspace{0.2cm}  \sum_{n=1}^{N_s}B_{kA}\left[R_{kA}(\pmb{\Gamma}[n]) -\hat{R}_{kj}(\pmb{\Gamma}[n]; \pmb{\Gamma}^{(z)}[n])\right]^+ \ge s_k, \nonumber\\
&& \hspace{3.2cm}\text{for} \hspace{0.2cm} k,j \in \mathcal{K}  \hspace{0.2cm} \text{and} \hspace{0.2cm} j \neq k, \label{eq:upConst3}\\
&& \hspace{-0.8cm}\text{(\ref{eq:s2}), (\ref{eq:pwrconst}), (\ref{eq:rhoconst}), (\ref{eq:avgmaxpwr})},
\end{eqnarray}
\end{subequations}
where $R_{kA}(\pmb{\Gamma}[n]) \triangleq R_{kA}(\pi_k[n], \pmb{p}_U[n], \pmb{\Theta}[n])$. Here, the problem (\ref{eq:pwropt2}) is convex, and thus has a unique solution $\pmb{\Gamma}^*(\pmb{\Gamma}^{(z)})$ with $\pmb{\Gamma}^{(z)} = \{\pmb{\Gamma}^{(z)}[n]\}_{\forall n}$, which can be numerically obtained by standard convex optimization strategies or by toolboxes such as CVX MOSEK. Using the approximated problem (\ref{eq:pwropt2}) and SCA approach, we develop Algorithm \ref{al2} to achieve the iterative suboptimal solution of problem (\ref{eq:pwropt}).
\begin{algorithm} [t!]
\begin{algorithmic}
\caption{Joint power and offloading ratio optimization} \label{al2}
\State {\textbf{Input}}: $\pmb{\Gamma}^{(0)} = (\pmb{\pi}^{(0)})\in \pmb{\Omega}$ and the fixed $\pmb{p}_U$, $\pmb{\Theta}$. Set $z=0$.
\State {\textbf{Repeat} until a convergence criterion is satisfied}
\State \hspace{+0.2cm} Calculate $(\pmb{\Gamma}^*(\pmb{\Gamma}^{(z)}), \pmb{\rho}^*, \pmb{s}^*)$ using (\ref{eq:pwropt2}).
\State \hspace{+0.2cm} Set $\pmb{\Gamma}^{(z+1)}=\pmb{\Gamma}^{(z)}+\alpha^{(z)}(\pmb{\Gamma}^*(\pmb{\Gamma}^{(z)})-\pmb{\Gamma}^{(z)}), \hspace{+0.1cm} \alpha^{(z)} \in (0, 1]$.
\State \hspace{+0.2cm} Update $z \gets z+1$
\State {\textbf{Output}}: $\pmb{\Gamma}^*(\pmb{\Gamma}^{(z)}) = \pmb{\pi}^*, \pmb{\rho}^*, \pmb{s}^*$.
\end{algorithmic}
\end{algorithm}
Based on the discussions in Sec. \ref{sec:phaseTOpt} and Sec. \ref{sec:powOpt}, Algorithm \ref{al3} summarizes the proposed algorithm to obtain the solution to the problem (\ref{eq:opt1}). The fact that the total energy consumption of ground user devices is nonincreasing in every iteration ensures the convergence of Algorithm \ref{al3}. Specifically, Algorithm \ref{al3} comprises Algorithm \ref{al1} and Algorithm \ref{al2}, and their convergence is verified through \cite[Th. 2]{Scutari14Arxiv} when the sequence of the step size $\{\alpha^{(z)}\}$ in both algorithms is chosen as $\alpha^{(z)} \in (0, 1], \alpha^{(z)} \to 0$ and $\sum_{z}\alpha^{(z)}=\infty$.
In other words, $\{\alpha^{(z)}\}$ is a bounded sequence, and any of its limit points that are within $\pmb{\Gamma}^{(\infty)}$ corresponds to stationary points for problems (\ref{eq:phaseTopt3}) and (\ref{eq:pwropt2}). Moreover, the non-existence of the local minimum for the original problem (\ref{eq:opt2}) occurs if Algorithm \ref{al3} continues indefinitely, as there is no limit point $\pmb{\Gamma}^{(\infty)}$ in (\ref{eq:phaseTopt3}) and (\ref{eq:pwropt2}).

\begin{algorithm} [t!]
\begin{algorithmic}
\caption{Proposed algorithm for energy-efficient secure offloading System designed via UAV-mounted IRS} \label{al3}
\State {\textbf{Initialize}}: $\pmb{\Gamma}^{(0)} = (\pmb{p}^{(0)}_U, \pmb{q}^{(0)}, \pmb{r}^{(0)}, \pmb{\pi}^{(0)})\in \pmb{\Omega}$ and the fixed $\pmb{s}, \pmb{\rho}$. Set $z=0$.
\State {\textbf{Repeat} until a convergence criterion is satisfied}
\State \hspace{+0.2cm} With given ($\pmb{\Gamma}^{(z)}, \pmb{s}$), obtain ($\pmb{p}_U^{(z+1)}, \pmb{q}^{(z+1)}, \pmb{r}^{(z+1)}, \pmb{\Theta}^{*}$)
\State \hspace{+0.2cm} using Algorithm \ref{al1}.
\State \hspace{+0.2cm} With given $(\pmb{p}_U^{(z+1)}, \pmb{\Theta}^{*})$, obtain ($\pmb{\pi}^{(z+1)}, \pmb{s}^{*}, \pmb{\rho}^{*}$)
\State \hspace{+0.2cm} using Algorithm \ref{al2}.
\State \hspace{+0.2cm} Update $z \gets z+1$
\State {\textbf{Output}}: $\pmb{\pi}^*, \pmb{p}_U^*, \pmb{\Theta}^*, \pmb{\rho}^*$.
\end{algorithmic}
\end{algorithm}
\section{Propulsion Energy Model of UAV-mounted IRS}\label{sec:opt2}
In Sec. \ref{sec:flyingE}, we consider the flying energy model of UAV-mounted IRS in (\ref{eq:flyingE}), which relies only on the velocity of UAV-mounted IRS. Here, we follow the more elaborate aerodynamic model for flying energy consumption \cite{Leishman06, Filippone06, JSA18TVT}, based on which we develop the energy-efficient secure offloading techniques for the better understanding about the effect of the flying energy model on the system design.

First, we define the acceleration of UAV-mounted IRS as $\pmb{a}_U[n]=(\pmb{v}_U[n+1]-\pmb{v}_U[n])/t_s$. Since the velocity $\pmb{v}_U[n]$ and acceleration $\pmb{a}_U[n]$ are dependent on the positions of UAV-mounted IRS $\{\pmb{p}_U[n]\}_{\forall n}$, we can express their relation based on the quadratic Taylor approximation model as follows:
\begin{equation}\label{eq:pva}
\pmb{p}_U[n+1]=\pmb{p}_U[n]+\pmb{v}_U[n]t_s + \frac{1}{2}\pmb{a}_U[n]t_s^2.
\end{equation}
By following \cite{Leishman06, Filippone06, JSA18TVT}, the flying energy model for UAV-mounted IRS at each slot $n$ can be then modeled as follows:
\begin{eqnarray}\label{eq:flyingE2}
&&\hspace{-0.9cm}E_U^{\text{F}}(\pmb{v}_U[n], \pmb{a}_U[n]) = \kappa_1\left\|\pmb{v}_U[n]\right\|^3 \nonumber\\
&&\hspace{2.2cm} + \frac{\kappa_2}{\left\|\pmb{v}_U[n]\right\|}\left(1
+\frac{\left\|\pmb{a}_U[n]\right\|^2}{g^2}\right),
\end{eqnarray}
where $g$ indicates gravitational acceleration, and the constants $\kappa_1$ and $\kappa_2$ represent the UAV operational capability such as drag coefficient, the area of main rotor disk, and induced power factor, the details of which can be found in \cite[Appendix C]{JSA18TVT}. Accordingly, considering (\ref{eq:flyingE2}) in lieu of (\ref{eq:flyingE}) in (\ref{eq:opt2}), the optimization problem (\ref{eq:opt2}) can be reformulated as
\begin{subequations}\label{eq:opt3}
\begin{eqnarray}
&&\hspace{-1.5cm} \underset{\substack{\pmb{\pi},\pmb{p}_U, \pmb{v}_U,\\
\pmb{a}_U, \pmb{\Theta}, \pmb{\rho}, \pmb{s}}}{\min} \hspace{0.2cm} \sum_{k=1}^K\sum_{n=1}^{N_s}E_k^{\text{Comm}}(\pi_k[n]) + \sum_{k=1}^K E^\text{Local}_k(\rho_kI_k) \label{eq:obj3}\\
&&\hspace{-1.5cm} \text{s.t.} \hspace{0.2cm} \sum_{n=1}^{N_s}E_{U}^{\text{F}}(\pmb{v}_U[n], \pmb{a}_U[n]) \le E_{th}, \label{eq:UAVConst3}\\
&&\hspace{-0.8cm} \pmb{v}_U[n+1] = \pmb{v}_U[n] + \pmb{a}_U[n]t_s, \hspace{0.2cm} \text{for} \hspace{0.2cm} n \in \mathcal{N}, \label{eq:va}
\end{eqnarray}
\begin{eqnarray}
&&\hspace{-3.0cm} \pmb{v}_U[1] = \pmb{v}_U[N_s+1] = \pmb{v}_U, \label{eq:vifConst}\\
&&\hspace{-3.0cm} \pmb{a}_U[n] \le \pmb{a}_U^{\max}, \hspace{0.2cm} \text{for} \hspace{0.2cm} n \in \mathcal{N}, \label{eq:amax}\\
&&\hspace{-3.0cm}  \text{(\ref{eq:pva}), (\ref{eq:s2}), (\ref{eq:upConst2}),  (\ref{eq:posConst}) - (\ref{eq:opt1remaining})},\label{eq:opt3remaining}
\end{eqnarray}
\end{subequations}
where $\pmb{a}_U=\{\pmb{a}_U[n]\}_{\forall n}$, (\ref{eq:vifConst}) is the constraint about the initial and terminal UAV-mounted IRS velocity, and (\ref{eq:amax}) is the constraint for the maximum acceleration of UAV-mounted IRS. Compared to (\ref{eq:opt2}), the additional optimization variables $\pmb{a}_U$ are considered in (\ref{eq:opt3}) by adopting the new flying energy model (\ref{eq:flyingE2}) of UAV-mounted IRS, which yields the modified or additional constraints (\ref{eq:UAVConst3}) - (\ref{eq:amax}) and (\ref{eq:pva}). Among them, the modified constraint (\ref{eq:UAVConst3}) is the only nonconvex constraint, while the other newly-added constraints are convex. To address the nonconvexity of (\ref{eq:UAVConst3}), we introduce the slack variable $\nu_n$, while satisfying the condition $0 \le \nu_n \le \|\pmb{v}_U[n]\|$ for all $n$. Accordingly, the constraint (\ref{eq:UAVConst3}) is upper-bounded as
\begin{eqnarray}\label{eq:flyingE2_up}
&&\hspace{-1.2cm}E_{U}^{\text{F}}(\pmb{v}_U[n], \pmb{a}_U[n]) \le \kappa_1\left\|\pmb{v}_U[n]\right\|^3 + \frac{\kappa_2}{\nu_n} + \frac{\kappa_2\left\|\pmb{a}_U[n]\right\|^2}{g^2\nu_n} \nonumber\\
&&\hspace{-0.9cm}\overset{(a)}{\le} \kappa_1\left\|\pmb{v}_U[n]\right\|^3 + \frac{\kappa_2}{\nu_n} + \frac{\kappa_2}{2g^2}\left(\left\|\pmb{a}_U[n]\right\|^2 + \frac{1}{\nu_n}\right)^2 \nonumber\\
&&\hspace{-0.6cm} - \frac{\kappa_2}{2g^2}\left(\left\|\pmb{a}_U^{(z)}[n]\right\|^4 + \frac{1}{\left(\nu_n^{(z)}\right)^2}\right)\nonumber\\
&&\hspace{-0.6cm} -\frac{\kappa_2}{g^2}\left(2\left\|\pmb{a}_U^{(z)}[n]\right\|^2\left(\pmb{a}_U^{(z)}[n]\right)^T\left(\pmb{a}_U[n]-\pmb{a}_U^{(z)}[n]\right)\right.\nonumber\\
&&\hspace{-0.6cm}-\frac{1}{\left(\nu_n^{(z)}\right)^3}\left(\nu_n-\nu_n^{(z)}\right) \bigg) \nonumber\\
&&\hspace{-0.9cm} \triangleq \hat{E}_{U}^{\text{F}}(\pmb{\Gamma}[n], \pmb{\Gamma}^{(z)}[n]),
\end{eqnarray}
where (a) can be obtained by Lemma \ref{lem2}, and we redefine the set of variables $\pmb{\Gamma}$ by including the additional variables $\pmb{a}_U=\{\pmb{a}_U[n]\}_{\forall n}$ and $\pmb{\nu}=\{\nu_n\}_{\forall n}$ as $\pmb{\Gamma}=\{\pmb{\pi}, \pmb{p}_U, \pmb{v}_U,\pmb{a}_U, \pmb{\Theta}, \pmb{\rho}, \pmb{s}, \pmb{\nu}\}$ and $\pmb{\Gamma}^{(z)}$ as the related set for the $z$th iterate of the SCA algorithm. Then, the problem (\ref{eq:opt3}) can be reformulated by considering (\ref{eq:flyingE2_up}) for (\ref{eq:UAVConst3}) given as
\begin{subequations}\label{eq:opt4}
\begin{eqnarray}
&&\hspace{-2.1cm} \min_{\pmb{\Gamma}} \hspace{0.2cm} \sum_{k=1}^K\sum_{n=1}^{N_s}E_k^{\text{Comm}}(\pmb{\Gamma}[n]) + \sum_{k=1}^K E^\text{Local}_k(\pmb{\Gamma}) \label{eq:obj4}\\
&&\hspace{-2.1cm} \text{s.t.} \hspace{0.2cm} \sum_{n=1}^{N_s}\hat{E}_{U}^{\text{F}}(\pmb{\Gamma}[n], \pmb{\Gamma}^{(z)}[n]) \le E_{th}, \label{eq:UAVConst3_bounded}\\
&&\hspace{-1.4cm} \nu_n^2 \le f^{LB}(\pmb{\Gamma}[n], \pmb{\Gamma}^{(z)}[n]), \hspace{0.2cm} \text{for} \hspace{0.2cm} n \in \mathcal{N}, \label{eq:nu}\\
&&\hspace{-1.4cm} \nu_n \ge 0, \hspace{0.2cm} \text{for} \hspace{0.2cm} n \in \mathcal{N}, \label{eq:nupos}\\
&&\hspace{-1.4cm}  \text{(\ref{eq:va}) - (\ref{eq:opt3remaining})},
\end{eqnarray}
\end{subequations}
where the squared norm $\|\pmb{v}_U[n]\|^2$ is limited by the linear lower bound $\le f^{LB}(\pmb{\Gamma}[n], \pmb{\Gamma}^{(z)}[n])$ as
\begin{eqnarray}
\label{eq:flb}
&&\hspace{-1.2cm} f^{LB}(\pmb{\Gamma}[n], \pmb{\Gamma}^{(z)}[n]) = \|\pmb{v}_U^{(z)}[n]\|^2 \nonumber\\
 &&\hspace{0.3cm} + 2\|\pmb{v}_U^{(z)}[n]\|^T(\pmb{v}_U[n]-\pmb{v}_U^{(z)}[n]) \le \|\pmb{v}_U[n]\|^2.
\end{eqnarray}
The problem (\ref{eq:opt4}) can be handled with the proposed Algorithm.

\begin{table}[t!]
    \renewcommand\arraystretch{1.5}
    \caption{Simulation parameters} \label{sim:parameter} 
    \resizebox{\columnwidth}{!}{%
    \begin{tabular}{ c c c c c c } 
    \hline\hline 
    Parameter & Value & Parameter & Value & Parameter & Value\\
    \hline 
   
    $B_{kA}$ & 0.25 MHz & $t_s$ & 1 s & $N_0$ & -174 dBm/Hz \\
    $v_{\max}$ & 10 m/s & $a_{\max}$ & 5 $\text{m}/\text{s}^2$ &  $m_u$ & 9.75 kg \\
    $g$ & 9.8 $\text{m}/\text{s}^2$ & $\kappa_1$ & 0.0822 & $\kappa_2$ & 111.57 \\
    $H$ & 90 m & $E_{th}$ & 20 kJ & $C_k$ & 1550.7\\
    $\gamma_k$ & $10^{-26}$ & $d/\lambda$ & 0.5 & $g_0$ & -35 dB \\
    $P_{T}^{\text{avg}}$ & 30 dBm & $ P_{T}^{\text{max}} $ & 40 dBm & - & - \\
    \hline\hline
    \label{parameter_table} \end{tabular}}
 \end{table}

\section{Numerical Results}\label{sec:num}

\begin{figure}[t]
\centering
\includegraphics[width=8.5cm]{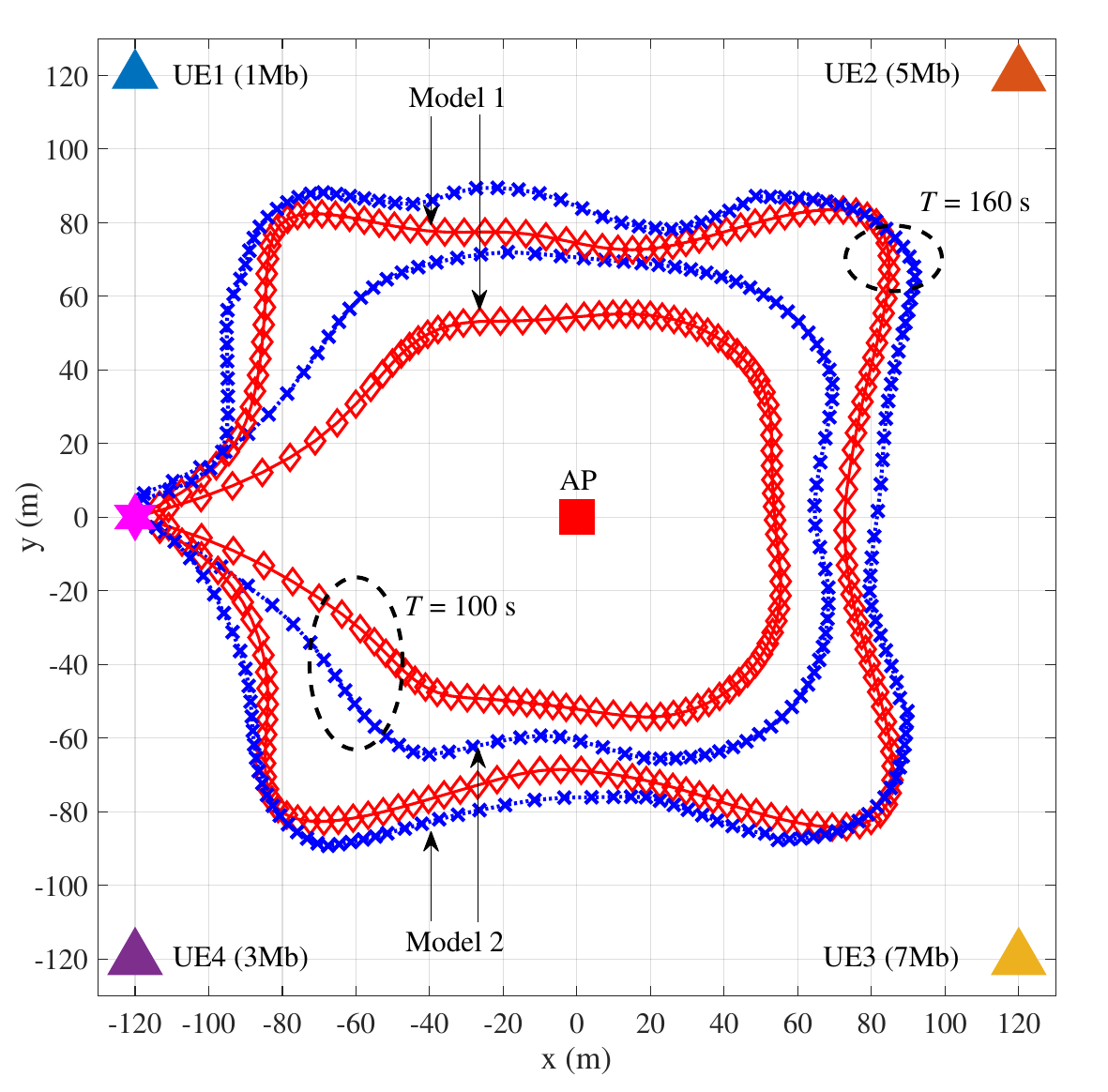}
\caption{Optimized trajectory of proposed algorithm with a mission time of $T$ = 100 s and $T$ = 160 s ($K=4$, $(I_1, I_2, I_3, I_4)=(1, 5, 7, 3)$ Mbits, $\pmb{p}_1=(-120, 120, 0)$ m, $\pmb{p}_2=(120, 120, 0)$ m, $\pmb{p}_3=(120, -120, 0)$ m, $\pmb{p}_4=(-120, -120, 0)$ m, $\pmb{p}_A=(0, 0, 0)$ m, $\pmb{p}_U[1]=\pmb{p}_U[N_s+1]=(-120, 0)$, $H=90$m and $L=16$).}
\label{fig:sim_1}
\end{figure}

In this section, we verify the performance of the proposed algorithm by jointly optimizing the transmit power of users, offloading ratio, phase shift matrix and path planning of UAV-mounted IRS via several numerical results. We adopt the flying models given in Sections \ref{sec:opt} and \ref{sec:opt2}, where the flying energy model is developed by (\ref{eq:flyingE}) (Model 1) or (\ref{eq:flyingE2}) (Model 2). For highlighting the superiority of the proposed algorithm, we compare the following reference schemes: \textit{(i) No trajectory optimization and identity phase} (No traj.opt. \& identity phase): In this scheme, we assume that the phase shift introduced by the passive elements can be represented as a identity matrix, which implies that the phase relation between users and the AP during the uplink communication remains invariant. \textit{(ii) No trajectory optimization and fixed phase} (No traj.opt. \& fixed phase): In this scheme, the phase shift of the passive reflecting elements is fixed with the certain directions, which is established by dividing the mission area into several sections by connecting the AP and the midpoints of two adjacent users, and the phase is fixed for the direction of the divided area when the UAV-mounted IRS is on between the user and the AP. \textit{(iii) No trajectory optimization and optimal phase} (No traj. opt. \& opt. phase): With this scheme, the UAV-mounted IRS with the optimized phase flies along the fixed trajectory for the users, whose transmit power and offloading ratio are optimized by the proposed algorithm. Unless specified otherwise, the parameters detailed in Table \ref{sim:parameter} are used in simulations. The initial and terminal velocities of the UAV-mounted IRS are set to $\pmb{v}_U = \lVert\pmb{v}_U\rVert(\pmb{p}_U[N_s+1]-\pmb{p}_U[1])/\lVert \pmb{p}_U[N_s+1]-\pmb{p}_U[1]\rVert$, where $\lVert\pmb{v}_U\rVert \le v_{\max}$, and $\kappa_1$ and $\kappa_2$ are calculated by considering the fixed-wing UAV-mounted IRS according to \cite{JSA18TVT} for Model 2.
\begin{figure}[t!]
\centering
\includegraphics[width=8.5cm]{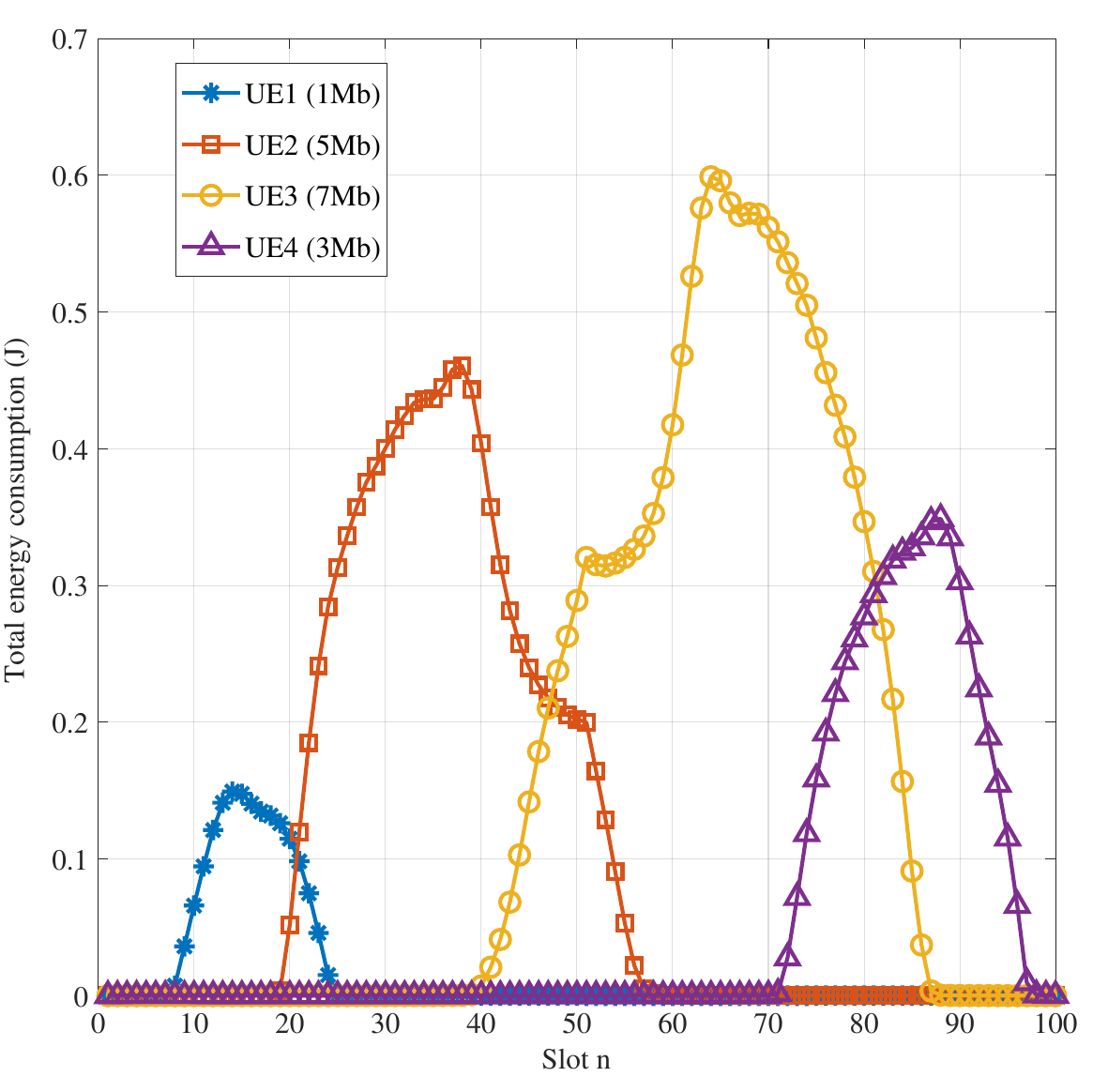}
\caption{Uplink transmit power allocation of ground users under Model 1 of Fig. \ref{fig:sim_1}.}
\label{fig:sim_2}
\end{figure}

As shown in Fig. \ref{fig:sim_1}, we consider 240 m $\times$ 240 m square area with $K=4$ users in each corner of the square as $\pmb{p}_1=(-120, 120, 0)$, $\pmb{p}_2=(120, 120, 0)$, $\pmb{p}_3=(120, -120, 0)$ and $\pmb{p}_4=(-120, -120, 0)$ respectively, and AP at the center as $\pmb{p}_A=(0, 0, 0)$, while the initial and terminal coordinates of UAV-mounted IRS are $\pmb{p}_U[1]=\pmb{p}_U[N_s+1]=(-120, 0)$ with initial speed $\lVert \pmb{v}_U \rVert = 7.54 \ \text{m}/\text{s}$ for Model 2. The number of bits required to be processed by each ground user is set to be $I_1=1$ Mbits, $I_2=5$ Mbits, $I_3=7$ Mbits, and $I_3=3$ Mbits, respectively, and the number of passive reflecting elements is fixed as $L = 16$. We investigated the trajectories under different mission times $T$ = 100 s and $T$ = 160 s, in both flying models with fixed altitude $H$ = 90 m.

\begin{figure}[t]
\centering
\includegraphics[width=8.5cm]{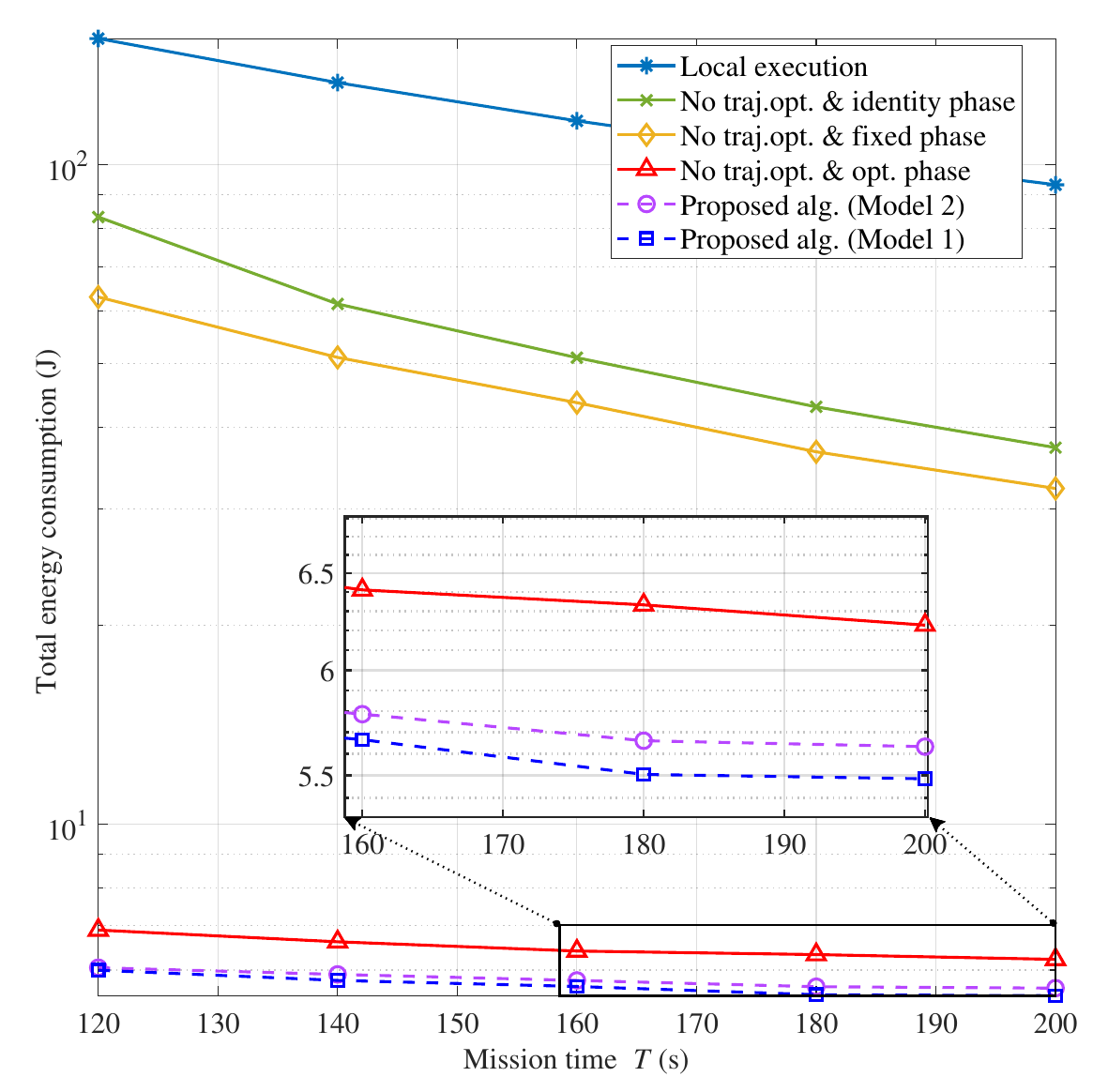}
\caption{Total energy consumption of ground users versus the mission time $T$ ($K=4$, $(I_1, I_2, I_3, I_4)=(5, 5, 5, 5)$ Mbits, $\pmb{p}_1=(-90, 90, 0)$ m, $\pmb{p}_2=(90, 90, 0)$ m, $\pmb{p}_3=(90, -90, 0)$ m, $\pmb{p}_4=(-90, -90, 0)$ m, $\pmb{p}_A=(0, 0, 0)$ m, $\pmb{p}_U[1]=\pmb{p}_U[N_s+1]=(-90, 0)$, $L=16$ and $H=90$m).}
\label{fig:sim_3}
\end{figure}

In Fig. \ref{fig:sim_1}, the optimized trajectories are illustrated with different mission times according to both flying energy models. Under both models, we observe that the UAV-mounted IRS stays longer in the region between each user and the AP. Moreover, it is shown in Fig. \ref{fig:sim_1} that the more bits the user offloads, the longer the UAV-mounted IRS tends to stay for fulfilling the secure offloading requirements. Within the sufficient mission time of 160 s, the UAV-mounted IRS travels to be closer to each user to avoid internal eavesdropping of other users. This tendency decreases the total energy consumption of ground user devices as the mission time is increased (see, Fig. \ref{fig:sim_3}). Although the above observations are attained under both flying energy Models, Model 2 tends to achieve the smoother trajectory compared to Model 1 owing to the energy consumption defined in terms of accelerations. 
In Model 1 of Fig. \ref{fig:sim_1} with $T$ = 100 s, the optimized power allocation of ground users is depicted in Fig. \ref{fig:sim_2}.
It is observed that the higher power and the more time slots are allocated to the users with more bits to be processed. Moreover, the power allocation of each user is concentrated especially when the UAV-mounted IRS is nearby to guarantee the secrecy rate constraint. This aspect can be seen regardless of the type of flying energy model and mission time (not reported here).

In Fig. \ref{fig:sim_3}, we analyze the total energy consumption of ground user devices of the proposed algorithm as a function of the mission time $T$. We execute the experiment in the $180\times180$ $\text{m}^2$ square area with $K=4$ UEs, each with the same input bits as 5 Mbits, deployed on the edges and the AP in the center. We set the initial and terminal coordinates of UAV-mounted IRS as $\pmb{p}_U[1]=\pmb{p}_U[N_s+1]=(-90, 0)$, and Table \ref{sim:parameter} covers the rest of the parameters that are not mentioned. As seen in Fig. \ref{fig:sim_3}, the proposed algorithm of jointly optimizing user transmit power, offloading ratio, UAV-mounted IRS phase shift and trajectory achieves the considerable gains in terms of total energy consumption. Although there is no substantial difference between the UAV-mounted IRS flying energy Model 1 (5.51 J at $T$ = 180 s) and Model 2 (5.68 J at $T$ = 180 s) with the proposed scheme, it is clear that using Model 1 induces the less total energy consumption of ground users than Model 2. This is because the rotary wing model is more flexible than the propulsion wing-based model in positioning to the optimal position. Notably, the use of fixed phase with the predefined directions attains the 17.03\% decrease compared to the identity phase, while the optimal phase offers the 89.23\% reduction in the total energy consumption compared to the identity phase, indicating the important role of the phase shift ability of the UAV-mounted IRS as a key performance determinant.

\begin{figure}[t]
\centering
\includegraphics[width=8.5cm]{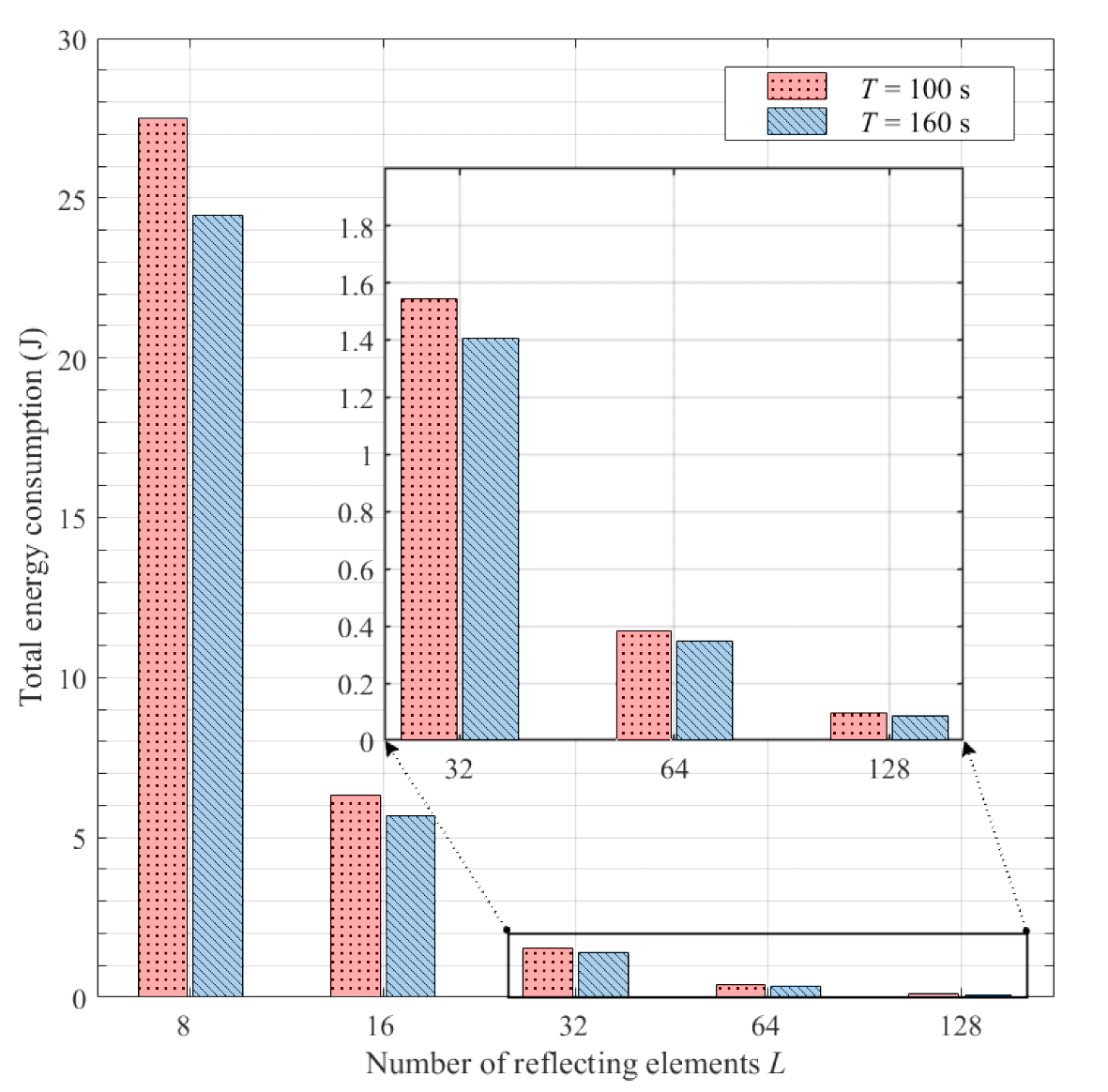}
\caption{Total energy consumption of ground users versus the number of reflecting elements $L$ in the same setting of Fig. \ref{fig:sim_3} with Model 1.}
\label{fig:sim_4}
\end{figure}

Fig. \ref{fig:sim_4} illustrates the total energy consumption of ground users versus the number of UAV-mounted IRS reflecting elements $L$ under the same scenario of Fig. \ref{fig:sim_3}, with Model 1. With the increasing number of reflecting elements, the superiority of the proposed algorithm is more emphasized in the total energy consumption of ground users. For instance, when the mission time is $T$ = 100 s, the energy consumption with $L$ = 128 is 0.09J, which is only 1.5\% of the energy consumed with $L$ = 16. This can be explained that the passive reflecting architecture of UAV-mounted IRS can allow to provide the additional spatial multiplexing gain without the specific energy. Furthermore, as the number of reflecting elements increases, the difference in energy consumption between mission times of 100 s and 160 s decreases, indicating that the increase in reflecting elements resulted in the smaller gap in the energy consumption of users between different mission times.

\section{Conclusions}\label{sec:con}
In this paper, we have investigated an energy-efficient secure offloading system with the assistance of UAV-mounted IRS. To this end, we aim to minimize the total energy consumption of ground users by jointly optimizing their transmit power and offloading ratio as well as the trajectory and phase shift of UAV-mounted IRS. We have considered the constraints for secure offloading accomplishment and the energy budget and mobility of UAV-mounted IRS based on different flying energy models. We have proposed algorithm based on SCA method, whose superiority is verified via simulations. Our numerical results have shown the considerable energy savings achieved by our proposed algorithm compared to local execution and partial optimizations, highlighting the importance of the phase shift design and path planning of the UAV-mounted IRS in achieving energy-efficient secure offloading. As our future works, the multiple UAV-mounted IRSs with the different height and the terrestrial IRSs can be explored for secure offloading system.


\bibliographystyle{unsrt}
\bibliographystyle{IEEEtran}
\bibliography{UAV_mounted_IRS_Kim}
\end{document}